\documentclass{article}

\begin{document}

\title{Stability of an electron embedded in Higgs condensate}         
\author{Eugen \v{S}im\'{a}nek  \footnote {electronic address: simanek@ucr.edu} \\Department of Physics, University of California, Riverside, Calif. 92521}  

\date{}          
\maketitle

\begin{abstract}

We study stability of an electron distributed on the surface of a spherical cavity in Higgs condensate.  The surface tension of the cavity prevents the electron from flying apart due to Coulomb repulsion.  A similar model was introduced by Dirac in 1962, though without reference to Higgs condensate.  In his model, the equilibrium radius of the electron equals the classical electron radius, $R^{c}_{e} \simeq 2.8  \times 10^{-13}$ cm, that is about $10^{5}$ times the radius consistent with experimental data.  To address this problem, we replace the Coulomb term in the total energy of the electron by fermion self-energy involving screening by electrons occupying the negative energies of the vacuum.  The tension of the cavity is obtained using the approximation $\xi_{0} \ll R_{0}$ where $\xi_{0}$ is the coherence length.  For $\xi_{0} = 10^{-3} R_{0}$, the equilibrium radius in this model is $R_{0} \simeq 9.2 \times 10^{-32}$ cm.  

For such a small radius, we find the gravitational energy of the electron to be large enough to cancel the energy $\hbar c/R$, coming from the vibrational zero point energy and the kinetic energy of the embedded electron.
\end{abstract}

PACS number(s):  12.15.-y, 12.20.-m, 12.39.Ba, 14.60.Cd, 14.80.Bn

\section{Introduction}

The problem of stability of an electron has an old history.  It began with the introduction of the concept of an electron of finite size first proposed by Abraham [1] and Lorentz [2].  This concept is, however, faced with difficulties.  The main problem comes from the assumption that the mass of the electron is entirely of electromagnetic origin and that no non-electromagnetic mass exists [1,2].  First, it is clear that a finite charge distribution cannot be stable under pure electromagnetic forces.  There is also a problem with the relativistic transformation properties of the energy and momentum of the electromagnetic field of the electron since these quantities do not form a 4-vector.  These difficulties can only be overcome by introducing non-electromagnetic stresses [3, 4].

Dirac [5] introduced in 1962 a simple model of an electron which consists of a charged conducting surface of a cavity in the electromagnetic field.  The surface tension of the cavity prevents the electron from flying apart under the Coulomb repulsion of its surface charge.  This idea offers a simple phenomenological solution to the problem of the stability of electron.  However, the origin of the surface tension is not addressed.  Using this model, Dirac [5], attempts to develop a theory of the muon by allowing the radius of the cavity to oscillate about its equilibrium value. By associating the lowest excited state with the muon, Dirac makes estimates of the mass of the muon.

In the present paper, we study a model similar to that of Dirac, however, we confine ourselves to the problem of electron stability. First, we focus on the origin of surface tension.  We note that Dirac's idea [5] of leptons has been transferred to hadrons where it formed the basis of the bag model [6], [7]. According to Ref. [7], the hadron bag is situated in a cavity that is formed in space filled with a condensate of scalar bosons (Higgs condensate).  The cavity exhibits a surface tension caused by the sudden drop of the condensate density at the interior of the cavity [8].  

We assume that the surface tension of the cavity surrounding the electron charge has a similar origin.  This enables us to derive an expression for the surface tension that depends on the parameters defining the Higgs potential in the electroweak gauge-model theory [9-11]. With use of this expression, we establish a formula for the total energy of the system consisting of the electron charge and the surrounding cavity.  This formula involves the vacuum expectation value of the Higgs field which is known and the Ginzburg-Landau coherence length [12] which, however, depends on parameters of the Higgs potential that are not known [11].  

We propose a following departure from the Dirac model [5].  The Coulomb self-energy term, in the equation for the total energy, is replaced by fermion self-energy which takes into account the presence of occupied negative energy levels in the vacuum.  This type of self-energy was first derived by Weisskopf [13]. It provides a strong screening of the Coulomb energy that is essential for bringing the total energy closer to the actual value known for the rest energy of an electron.

The surface tension of the caviy is evaluated in the approximation, $\xi_{0} \ll R_{0}$, where $\xi_{0}$ is the coherence length, and $R_{0}$ is the cavity radius (see Secs. 3-5).

A quantity that is of great importance for the theory of electron stability is the self-stress.  It was pointed out by Poincar\'{e} [3], that in order that the energy and momentum of the electron transform like a 4-vector, the self-stress must be equal to zero (see also Ref. [4]). 

In the present work, we show that the self-stress can vanish for a system consisting of the electron charge and the associated cavity owing to the fact that the nonvanishing self-stress, due to the Coulomb field, is cancelled by the Higgs's self-stress.  For the Dirac model, we show in Sec. 6 that exact cancellation takes place when the coherence length of the Higgs condensate is much smaller than the equilibrium radius of the cavity.  

The present work is concluded by a discussion of the role of short-range gravitational energy in justifying the simple form of total energy of Sec. 7.

\section{Lagrangian and Stress-Energy Tensor}       
Following the analogy to bag model of hadrons [7, 8], we consider a simple soliton model consisting of the Higgs field and the electric field due to the electron charge.  This charge is assumed to be uniformly spread on surface of a spherical cavity immersed in a vacuum filled with the Higgs condensate.

	The Lagrangian density for the model is written as

\begin{equation}\label{Eq1}
\pounds = \pounds^{H} + \pounds^{E}
\end{equation}

where $\pounds^{H}$is the contribution of the Higgs condensate

\begin{equation}\label{Eq2}
\pounds^{H}= \frac{1}{2} g^{\mu\nu}\partial_{\mu} \phi \partial_{\nu} \phi - V (\phi)
\end{equation}

The space-time metric tensor $g^{\mu\nu}$ has signature (1, -1,-1,-1), $\phi (\vec {r})$ is the scalar field, and $V (\phi)$ has a form ensuring spontaneous symmetry breaking [11]

\begin{equation}\label{Eq3}
V (\phi) = - \frac{1}{2} \mu^{2} |\phi|^{2} + \frac{1}{4}\lambda |\phi|^{4}
\end{equation}

The contribution of the electric field to $ \pounds$ is given by [15]

\begin{equation}\label{Eq4}
\pounds ^{E} = - \frac{1}{16 \pi} F_{\alpha \beta} F ^{\alpha \beta}
\end{equation}

where $F_{\alpha \beta}$ is the field-strength tensor.

Consistent with equation (1), the stress-energy tensor is written as

\begin{equation}\label{Eq5}
T _{ij} = T^{H}_{ij}+ T^{E}_{ij}
\end{equation}

	Using the expression [16]

\begin{equation}\label{Eq6}
T_{ij} = \frac{2}{\sqrt{|g|}} \frac{\partial}{\partial g^{ij}}\Big (\sqrt{|g|} \pounds \Big)
\end{equation}

we obtain with use of Eq. (2)

\begin{eqnarray}\label{Eq7}
T^{H}_{ij} = \partial_{i} \phi \partial_{j} \phi - g _{ij} \pounds ^{H} = \partial_{i} \phi \partial_{j} \phi \nonumber \\* + \Big [ \frac {1} {2} \sum ^{3}_{\mu = 1} (\partial _{\mu} \phi)^{2} + V (\phi)\Big ] g_{ij}
\end{eqnarray}
 
where the second equality follows by assuming a time-independent scalar field $(\partial_{0} \phi = 0)$.

	For the electro-magnetic stress-energy tensor, $T^{E}_{ij}$, we use the symmetrical tensor [15]

\begin{equation}\label{Eq8}
T^{E}_{ij} = \frac{1}{4 \pi} \Big(g_{\beta j} F_{i\lambda} F ^{\lambda \beta} + \frac{1}{4} g_{ij} F_{\mu \lambda} F^{ \mu \lambda} \Big)
\end{equation}

In the absence of magnetic fields, the field-strength tensor is given by [15]

\begin{equation}\label{Eq9}
F_{\alpha \beta} = \pmatrix {0 & E _{x} & E_{y} & E_{z}\cr -E_{x} & 0 & 0 & 0 \cr - E_{y} & 0 & 0 & 0 \cr - E_{z} & 0 & 0 & 0}
\end{equation}

	Of special interest to us is the component $T^{E}_{00}$ giving the Coulomb energy density, and $T^{E}_{xx}$ whose volume integral defines the self-stress [4].  Using Eqs. (8) and (9), we have

\begin{equation}\label{Eq10}
T^{E} _{00}= \frac {1} {8 \pi} (E^{2}_{x} + E^{2}_{y} + E^{2}_{z})
\end{equation}

and

\begin{equation}\label{Eq11}
T^{E}_{xx} = \frac {1}{8 \pi} (E^{2}_{y} + E^{2}_{z} - E ^{2}_{x})
\end{equation}

\section {Transformation of Energy and Momentum}

	In this section we examine the condition under which the volume integrals $\int T_{00} d \tau$ and $T_{10} d \tau$ transform like the energy and momentum of a particle [4].

	We consider a Lorentz transformation for two frames, K and K', in relative motion along axis $x^{1}$

\begin{eqnarray}\label{Eq12}
x^{0}= \gamma  (x^{'0} + \beta x ^{'1}) \nonumber \\* x^{1} = \gamma ( x^{'1}+ \beta  x ^{'0}) \nonumber \\* x ^{2} = x ^{'2} \nonumber \\* x^{3} = x^{'3}  
\end{eqnarray} 

where

\begin{equation}\label{Eq13}
\gamma = (1 - \beta^{2})^{-\frac{1}{2}}, \beta = v/c
\end{equation}

and $v$ is the relative velocity of the frames K and K'.  We assume that the electron is at rest in frame K so that $T_{10} = 0$.  

	The transformation of the stress-energy tensor is described by

\begin{equation}\label{Eq14}
T'_{\mu \nu} = \frac{\partial x^{\gamma}}{\partial x'^{\mu}} \frac{\partial x^{\delta}}{\partial x^{'\nu}} T_{\gamma \delta}
\end{equation}

Using Eq. (12), Eq. (14) yields

\begin{equation}\label{Eq15}
T'_{00} = \gamma^{2} (T_{00} + \beta ^{2} T _{11}) 
\end{equation}

\begin{equation} \label{Eq16}
T'_{10} = \gamma^{2} \beta (T_{00} + T_{11})
\end{equation}\label{Eq16}

There is also a Lorentz contraction of the volume element, $d \tau = d x^{1} dx^{2} dx^{3}$, given by

\begin{equation}\label{Eq17}
d \tau' = dx^{'1} dx ^{'2} dx^{'3} = \frac{1}{\gamma} d\tau
\end{equation}

From Eqs. (15)-(17), we obtain

\begin{equation}\label{Eq18}
\int T'_{00} d\tau ' = \gamma \Big( \int T_{00}d\tau + \beta^{2}\int T_{11}d\tau \Big)
\end{equation}

\begin{equation}\label{Eq19}
\int T'_{10} d\tau' = \gamma \beta \Big( \int T_{00}d \tau + \int T_{11} d \tau \Big)
\end{equation}

The transformation formulae for the energy $E$, and the momentum of a particle $P_{x}$ are [4]

\begin{eqnarray}\label{Eq20}
 E' = \gamma E  \nonumber \\* P'_{x} = - \frac{1}{c} \gamma \beta E   
\end{eqnarray}

On comparing this result with Eqs. (18) and (19), we see that $\int T_{00} d \tau$ and $- \frac{1}{c} \int T_{10} d \tau$ transform like the energy and momentum of a particle if and only if

\begin{equation}\label{Eq21}
\int T _{11} d \tau = 0
\end{equation}

The integral $\int T _{11} d \tau$ is called the self-stress. In Sec. 6 we examine validity of the condition (21) for the model described by the Lagrangian density (1).

\section{Surface Tension for Spherical Cavity}

	We consider a spherical cavity in unbounded space filled with Higgs condensate.  Our goal is to derive an expression for the surface tension which is formed at the cavity surface owing to the fact that the condensate density vanishes at the interior of the cavity.

	Starting from the Lagrangian density of Eqs. (1) and (2), the Euler-Lagrange equation for an isotropic scalar field $\phi (\vec{r})$ is

\begin{equation}\label{Eq22}
\nabla^{2} \phi = \frac{d V (\phi)}{d \phi} 
\end{equation}

Using the Laplacian in spherical coordinates, we have

\begin{equation}\label{Eq23}
\nabla^{2} \phi = \frac{d^{2} \phi}{d r^{2}} + \frac {2}{r} \frac{d \phi} {d r}
\end{equation}

As pointed out by Lee [8] the second term on the right-hand side of this equation can be neglected if the coherence length $\xi \ll R$.

This can be verified as follows.  The boundary conditions for the scalar field are

\begin{eqnarray}\label{Eq24}
\phi (r = R) = 0  \nonumber \\* \phi (r \rightarrow \infty) = \eta  \nonumber \\* \frac{d \phi}{dr} (r \rightarrow \infty) = 0
\end{eqnarray}

The parameter $\eta$ is the value of the scalar field at the locus of minima of $V (\phi)$.  Using Eq. (3), this yields

\begin{equation}\label{Eq25}
\eta^{2} = \frac{\mu^{2}} {\lambda}
\end{equation}

	To determine the coherence length we define a dimensionless field $f(r)$ by writing the scalar field as

\begin{equation}\label{Eq26}
\phi (\vec{r})= \eta f (r)
\end{equation}

Introducing this definition into Eq. (22), we obtain with the help of Eq. (24)

\begin{equation}\label{Eq27}
\frac{1}{\lambda \eta^{2}} \nabla^{2} f = - f + f ^{3}
\end{equation}

This equation defines a characteristic length, the Ginzburg-Landau coherence length [12]

\begin{equation}\label{Eq28}
\xi = \Big(\frac{1}{\lambda \eta^{2}}\Big)^{\frac{1}{2}}
\end{equation}

From the boundary conditions (24), we see that $\phi (r)$ rises from zero to $\eta$ in the interval $R \leq r < R + \xi$.  Thus, $\frac{d \phi}{dr} \simeq \frac{\eta}{\xi}$ and $\frac {d^{2}\phi}{dr^2} \simeq \frac {\eta}{\xi^{2}}$ in that region.  Consequently, the ratio of the two terms in Eq. (23) is given by

\begin{equation}\label{Eq29}
\frac{d^{2}\phi}{d r^{2}}\Big{/} \Big(\frac{2}{r}\frac{d \phi}{dr}\Big)\simeq R/2 \xi
\end{equation}

Therefore, if $\xi \ll R$, Eq.(22) can be simplified to one-dimensional soliton form [8]

\begin{equation}\label{Eq30}
\frac {d^{2} \phi} {dr^{2}} = \frac{d V}{d \phi}
\end{equation}

The first integral of this equation yields

\begin{equation}\label{Eq31}
\frac{1}{2} \Big( \frac{d\phi}{dr}\Big)^{2} - V (\phi) = \textrm{constant}
\end{equation}

where the constant is found by applying the boundary conditions (24).

Since $\phi (r) \rightarrow \eta$ as $r \rightarrow \infty$, we have using Eqs. (3) and (25)

\begin{equation}\label{Eq32}
V \Big(\phi (r \rightarrow \infty)\Big)\rightarrow - \frac{\mu^{2}}{2} \eta^{2} + \frac{\lambda}{4} \eta ^{4}= - \frac{\mu^{4}}{4 \lambda}
\end{equation}

Using this result and the fact that $\frac{d \phi}{dr} \rightarrow 0$ as $r \rightarrow \infty$, we obtain from Eq. (31) that the constant $= \frac{\lambda \eta^{4}}{4}$.  Consequently, Eq. (31) can be written as

\begin{equation}\label{Eq33}
\Big ( \frac{d \phi} {d r}\Big) ^{2} = \frac{\lambda}{2} \Big ( \phi^{2} - \eta ^{2} \Big) ^{2}
\end{equation}

Integrating this equation with use of the boundary conditions (24), we have

\begin{equation}\label{Eq34}
\int^{\phi (r)} _{\phi (R)} \frac{d \phi}{\eta^{2}- \phi ^{2}} = \Big( \frac{\lambda}{2}\Big)^{\frac{1}{2}} \int ^{r}_{R} dr
\end{equation}

Solving Eq. (34) for $\phi (r)$, we have

\begin{equation}\label{Eq35}
\phi (r) = \eta \; \textrm{tanh} \frac{r-R}{\xi_{0}}
\end{equation}

where $\xi_{0} = \sqrt {2} \xi$, $\xi$ being the Ginzburg-Landau coherence length given in Eq. (28).

The surface tension emerges from calculation of the energy of the Higgs field [12].  Using Eq. (7), this energy is given by

\begin{equation}\label{Eq36}
\int T^{H}_{00} d \tau = 4 \pi \int^{\infty}_{R} \Big[ \frac{1}{2}\eta^{2} f ^{'2} + V (\phi)\Big] r^{2} dr
\end{equation}

where $f ' = \frac{df}{dr}$.  With a hind-sight of a divergence that appears in the evaluation of the volume integral of $V(\phi)$, we write Eq. (3) as follows

\begin{equation}\label{Eq37}
V (\phi) = \frac{\lambda}{4} \Big( \phi^{2} - \eta^{2}\Big)^{2} - \frac{\lambda \eta^{4}}{4}
\end{equation}

Explicit evaluation of the volume integral of $V (\phi)$ shows that the first term on the right-hand side of Eq. (37) yields a finite integral, whereas the second term that is constant, produces a divergent integral proportional to the volume of the unbounded space.  Incidentally, the resulting divergent negative energy is equal to the condensation energy of the Higgs condensate of bosons uniformly spread over the entire space.

	Proceeding in the spirit of the renormalization theories [16, 17], we claim that this divergent energy is never observed and should be substracted as a "counterterm" from $V (\phi)$.  Thus, we define a renormalized potential

\begin{equation}\label{Eq38}
V_{\textrm{ren}} (\phi) = V (\phi) + \frac{\lambda \eta^{4}}{4} = \frac{\lambda}{4} \Big( \phi^{2} - \eta^{2}\Big)^{2}
\end{equation}

A similar subtraction is known to be done in theory of superconductivity to obtain the energy of N-S wall [18].

To evaluate the integral (36), we introduce a new variable

\begin{equation}\label{Eq39}
X = \frac {r-R}{\xi _{0}}
\end{equation}

Using Eqs. (35), (36) and (38), we have

\begin{eqnarray}\label{Eq40}
4 \pi \int ^{\infty}_{R} \Big [ \frac{1}{2} \eta^{2} f^{'2} + V_{\textrm{ren}} (\phi) \Big ] r^{2} dr \nonumber \\* = \frac{4 \pi \eta^{2}} {\xi _{0}} \int ^{\infty}_{0} (\textrm{sech} X)^{4} (R + \xi_{o} X)^{2} d X \nonumber \\* = \frac{4 \pi \eta^{2}}{\xi_{0}} (R^{2} I_{0} + 2 \xi_{0} R I _{1} + \xi^{2}_{0} I _{2})
\end{eqnarray}

where

\begin{equation}\label{Eq41}
I_{0} = \int^{\infty} _{0} (\textrm{sech} X)^{4} dX = \frac{2}{3}
\end{equation}

\begin{equation}\label{Eq42}
I_{1}= \int ^{\infty}_{0} (\textrm {sech} X)^{4}X dX = - \frac{2}{3} \textrm{ln} \frac{1}{2} - \frac{1}{6} \simeq 0.3
\end{equation}

\begin{equation}\label{Eq43}
I_{2} = \int^{\infty}_{0} (\textrm{sech} X)^{4} X^{2}dX - \frac{1}{3} \Big(\frac{\pi^{2}}{6}- 1 \Big)
\simeq 0.2\end{equation}

Out of the three terms on the right-hand side of Eq. (40), only the first one has the form of surface energy.  Using Eqs. (41-43), we see that this term is a dominant one if $R \gg \xi_{0}$.  The surface tension $\sigma$ is defined as the surface energy per unit area.  From Eqs. (40) and (41), we obtain

\begin{equation}\label{Eq44}
\sigma = \frac{2 \eta^{2}}{3 \xi_{0}} = \frac{4}{3}\xi_{0}|\epsilon _{\textrm{cond}}|
\end{equation}

where $\epsilon_{\textrm{cond}}= - \frac{\lambda \eta^{4}}{4}$ is the condensation energy density which makes appearance in Eqs. (37) and (38).

\section {Stability and Higgs condensate}

	In this section, we consider the energy of our model as a function of the radius, $R$, of the cavity.  The stable state of the system is determined by the value of $R$ that minimizes the energy.  This is similar to Dirac's treatment of the extensible electron model [5].  

Starting from Eq. (10), the Coulomb energy of the electron charge distributed on the cavity surface is 

\begin{equation}\label{Eq45}
\int T^{E}_{00} d\tau = \frac{1}{2} e^{2} \int^{\infty}_{R} \frac{dr}{r^{2}} = \frac{e^{2}}{2R}
\end{equation}

Upon adding to this result the surface energy, the total energy, $E_{\textrm {tot}}$, becomes

\begin{equation}\label{Eq46}
E_{\textrm{tot}} (R) = \frac{e^{2}}{2R} + 4 \pi \sigma R^{2}
\end{equation}

The equilibrium radius, $R_{0}$, is found from the equation $\partial E_{ \textrm {tot}} / \partial R = 0$.  Using Eq. (46), this yields

\begin{equation}\label{Eq47}
R_{0} = \Big ( \frac{e^{2}}{16 \pi \sigma} \Big)^{\frac{1}{3}}
\end{equation}

Then the equilibrium value of Coulomb energy is given by

\begin{equation}\label{Eq48}
\frac{e^{2}}{2R_{0}} = \frac{1}{2} e ^{\frac{4}{3}} (16 \pi \sigma)^{\frac {1}{3}}
\end{equation}

and the surface energy at equilibrium is

\begin{equation}\label{Eq49}
4 \pi \sigma R^{2}_{0} = \frac{1}{4} e^{\frac{4}{3}}(16 \pi \sigma)^ \frac{1}{3}
\end{equation}

Hence, consistent with Dirac's paper [5], we find that the Coulomb part of $E_{\textrm{tot}}(R_{0})$ is twice as large as the surface one.  Consequently, the total energy becomes 

\begin{equation}\label{Eq50}
E_{\textrm{tot}} (R_{0})= \frac{3}{4}\frac{e^{2}}R_{0}
\end{equation}

In Sec. 7, we use this equation to relate the equilibrium radius of the cavity to the classical radius of the electron. 

\section{Proof of Eq. (21)}

In what follows, we establish the identiy (21) by proving that, as long as $\xi _{0} \ll R_{0}$, the self-stress due to the Coulomb field is cancelled by the Higgs self-stress.  Thus, we shall verify the identity

\begin{equation}\label{Eq51}
\int T^{E}_{xx} d\tau = - \int T^{H}_{xx} d\tau
\end{equation}

Using Eqs. (10) and (11) and invoking spatial isotropy, we have

\begin{equation}\label{Eq52}
\int T ^{E}_{xx} d\tau = \frac{1}{3} \int T ^{E}_{00} d\tau = \frac{e^{2}}{6 R_{0}}
\end{equation}

where, Eq. (45) is used to obtain the second equality.  Expressing $R_{0}$ with use of Eq. (47) and using Eq. (44) for $\sigma$, Eq. (52) yields

\begin{equation}\label{Eq53}
\int T^{E}_{xx} d\tau = \frac{16 \pi}{9}\frac{\eta^{2} R^{2}_{0}}{\xi_{0}}
\end{equation}

Setting $i = j = x$ in Eq. (7), we have

\begin{equation}\label{Eq54}
T^{H}_{xx} = \frac{1}{2} \Big [ (\partial _{x} \phi)^{2} - (\partial_{y} \phi)^{2} - (\partial_{z}\phi)^{2} \Big] - \frac {\lambda}{4}(\phi ^{2}- \eta^{2})^{2}
\end{equation}

Taking  account of spatial isotropy in the evaluation of the volume integrals of the first three terms, we obtain from Eq. (54)

\begin{equation}\label{Eq55}
\int T^{H}_{xx}d\tau = - \frac{1}{2} \int (\partial _{x}\phi)^{2} d\tau - \frac{\lambda}{4} \int (\phi^{2} - \eta^{2})^{2} d\tau
\end{equation}

From Eq. (26), we obtain using spherical coordinates, $\partial_{x} \phi = \eta f' (r) \textrm{sin} \theta \textrm {cos}\varphi$.  Consequently, we have

\begin{eqnarray}\label{Eq56}
\frac{1}{2} \int (\partial_{x}\phi)^{2} d\tau = \frac{1}{2}\eta^{2} \int ^{\infty}_{R_{0}} f^{'2}r^{2} dr \int ^{\pi}_{0} \textrm {sin}^{3} \theta d \theta \int^{2 \pi}_{0}\textrm{cos}^{2}\varphi d \varphi \nonumber \\ = \frac{2\pi}{3} \eta^{2}\int ^{\infty} _{R_{0}} f^{'2}r^{2}dr
\end{eqnarray}

The integral on the right-hand side of this equation has been evaluated in Eqs. (39-40).  For $R_{0} \gg \xi_{0}$, we obtain

\begin{equation}\label{Eq57}
\frac{1}{2} \int (\partial_{x}\phi)^{2} d \tau = \frac{4 \pi}{9} \frac{\eta^{2}R^{2}_{0}}{\xi_{0}}
\end{equation}

Using Eq. (26), the last term of Eq. (55) becomes

\begin{equation}\label{Eq58}
\frac{\lambda}{4}\int ^{\infty}_{R_{0}} (\phi^{2}- \eta^{2})^{2} d\tau = \pi \lambda \xi _{0}\eta^{4} \int^{\infty}_{0}(\textrm {tanh}^{2} X - 1)^{2}(R_{0}+ \xi_{0}X)^{2} dX
\end{equation}

where $X$ is the new variable defined in Eq. (39).  In the limit of $R_{0} \gg \xi_{0}$, Eq. (58) yields

\begin{equation}\label{Eq59}
\frac{\lambda}{4} \int ^{\infty}_{R}(\phi^{2} - \eta^{2})^{2} d\tau \simeq \frac{4 \pi}{3} \frac{\eta^{2}R^{2}_{0}}{\xi_{0}}
\end{equation}

Introducing Eqs. (57) and (59) into Eq. (55), we obtain

\begin{equation}\label{Eq60}
\int T^{H}_{xx} d\tau = - \frac{16 \pi}{9}\frac{\eta^{2} R^{2}_{0}}{\xi_{0}}
\end{equation}

Eqs. (53) and (60) imply that when $R_{0} \gg \xi_{0}$ the identity (51) holds so that the net self-stress vanishes

\begin{equation}\label{Eq61}
\int T_{11} d\tau = \int \Big ( T^{E}_{xx} + T ^{H}_{xx} \Big ) d\tau \simeq 0
\end{equation}

Hence, the condition (21) is verified.

\section {Fermion self-energy model}

According to Eq.(50), the rest mass of the classical electron is related to its radius in a way that is hard to reconcile with experiment.  For the electron mass given by $m \simeq 0.51$ MeV$/c^{2}$, this relation yields a radius $2.1 \times 10 ^{-13}$ cm.  However, experiments on electrons show no evidence of structure at the level of $10^{-18}$ cm. For instance, search for contact interaction at the LEP storage ring which probes electron structure at the 10 TeV range suggests electron radius $R_{0} < 2 \times 10 ^{-18}$ cm [14].

	In this section,  we show that this discrepancy can be possibly resolved if the Coulomb energy in Eq. (46) is replaced by fermion self-energy first obtained, with use of the Dirac hole theory, by Weisskopf [13].

Hence, instead of Eq. (46), we consider the following form for the total energy

\begin{equation}\label{Eq62}
E_{\textrm{tot}} = \frac{3 e^{2}mc}{2 \pi \hbar} \textrm{log} \Big ( \frac{\hbar}{mcR} \Big ) +  \frac{4 \pi}{3}R^{3} \Big(\frac{\eta^{2}}{2 \xi^{2}_{0}} \Big) 
\end{equation}

where the first term stems from the self-energy expression [4, 13] in which the cutoff wave-vector $k_ \textrm{max}$ is replaced by $1/R$.  

The second term represents the volume energy in the cavity with the energy density given by the magnitude of the condensation energy $V_{ren} (\phi = 0)$ (see Eq. (38)).  It turns out that this volume energy is of the order of $R_{0}/\xi_{0}$ larger than the surface energy $4 \pi \sigma R^{2}$.  Since we are assuming that $\xi_{0}/R_{0} \ll 1$, the surface energy term is not included in Eq. (62).
 
By setting $\partial E_\textrm{{tot}}/\partial R = 0$, the equilibrium radius is found to satisfy the following equation

\begin{equation}\label{Eq63}
- \frac{3 e^{2} mc} {2 \pi \hbar R_{0}} + \frac{2 \pi R^{2}_{0} \eta^{2}} {\xi^{2} _{0}} = 0
\end{equation}

From this equation, we have

\begin{equation}\label{Eq64}
R^{3}_{0} = \frac{3 e^{2}\xi^{2}_{0}mc^{2}}{4 \pi ^{2}\eta^{2} \hbar c}
\end{equation}

Using $mc^{2} = 0.51 $ MeV, $\eta^{2} \simeq 3 \times 10^{21}$ MeV/cm (see Ref. [11]), and noting that $e^{2}/\hbar c$ defines the fine-structure constant, $\alpha \simeq \frac {1}{137}$, we obtain from Eq. (64) $R^{3}_{0} \approx 9.2 \times 10 ^{-26} \xi^{2}_{0}$.  Consistent with the condition $\xi_{0} \ll R_{0}$, we substitute $\xi_{0} \approx 10^{-3} R_{0}$ into this relation and obtain

\begin{equation}\label{Eq65}
R_{0} \approx 9.2 \times 10 ^{-32} \textrm{cm}
\end{equation}

Using this result in Eq. (62), we estimate the predicted electron mass $\tilde{m} = E_ \textrm{tot}/c^{2}$.  With the parameters used on the right-hand side of Eq. (64), we obtain

\begin{equation}\label{Eq66}
E_ \textrm{tot} = A + B
\end{equation}

where $A \approx 0.15 m c^{2}$ and $B \approx 0.00116 m c^{2}$.  We see that the term $A$ which corresponds to the first term in Eq. (62), dominates the term $B$ representing the volume energy.  Thus, we conclude that, under the assumption of $\xi_{0} \approx 10 ^{-3} R_{0}$, Eq. (62) leads to the predicted electron mass $\tilde{m} \approx 0.15 m$.  We see that there is an 85 percent energy deficit that needs to be addressed. 

We conjecture that including Yukawa coupling between the fermion and the Higgs field may resolve this problem.  Following Ref. [11], the electroweak model leads to the following Yukawa coupling energy

\begin{equation}\label{Eq67}
E_{Y} = - \int d^{3} \vec{x} \pounds ^{e}_{Y} (\vec{x})
\end{equation}

where  $\pounds^{e}_{Y}$ is the Lagrangian density (see Eq. (4.52) of Ref. [11])

\begin{eqnarray}\label{Eq68}
 \pounds ^{e}_{Y} = - \frac{ G_{e}}{\sqrt {2}} \Big [ \phi (\vec{x}) (\bar{e}_{L} e _{R} + \bar{e}_{R} e_{L})  \Big ] \nonumber \\ = -  \frac{G_{e}} {\sqrt {2}} \Big [ \phi (\vec{x}) \bar{e} (\vec{x}) e (\vec{x}) \Big ]
\end{eqnarray}

Eqs. (67) and (68) imply that $E_{Y}$ is given by the overlap integral between the Higgs field and the electron density.  The magnitude of this overlap is not known, neither is the coupling constant $G_{e}$ (see Ref. [11]). Bardeen et al [7] consider a similar Yukawa term in their work on quark confinement.  They estimate that this term is much smaller than the kinetic energy $\hbar c/R$.  For the present problem of an embedded electron, the kinetic energy term tends to increase the quantity $\tilde {m}$ to a large value of order $10^{8}$ MeV/$c^{2}$ that is clearly incompatible with the electron mass $m= 0.5$ MeV$/c^{2}$ . In this context, we point out that the zero-point energy $\hbar \omega_{0}/2$ for the vibrations of the cavity radius is also of order $\hbar c/ R$ (see Ref. [19]).  

There is a way to deal with the above terms that are of order $\hbar c/R$.  First, we note that the equilibrium radius $R_{0}\approx 10 ^{-31}$cm, given in Eq. (65), is within the range of unification of all forces including gravity.  Garriga and Tanaka [20] consider the effect of Kaluza-Klein corrections to the metric of a spherically symmetric source of mass $M$ and find a modified Newtonian potential (see Eq. (23) of Ref. [20])

\begin{equation}\label{Eq69 }
\frac{h_{00}}{2}= \frac{GM}{r} \Big ( 1 + \frac {2 l^{2}}{3 r ^{2}} \Big )
\end{equation}

where $l$ is related to the cosmological constant in the bulk, $\Lambda$, by $\Lambda = - 6 l^{-2}$.  For $r\ll l$, the second term on the right hand side of (69) dominates the first one leading to the short-range gravitational potential of the form $h_{00}/ 2 \approx 2 GM l^{2}/3 r ^{3}$.  In Ref. [19], we calculate using this potential, the gravitational energy of a spherical electron mass of radius $R$ and obtain

\begin{equation}\label{Eq70}
E _{\textrm {grav}} = - \frac{2}{3} G_{N} \frac{m^{2}l^{2}}{R^{3}}
\end{equation}

where $m \approx 0.5$ MeV$/c^{2}$.  With $R_{0} \approx 10 ^{-31}$ cm, we obtain from Eq. (70), $E _{\textrm{grav}} \approx - 2 \times 10 ^{37} l^{2}$ MeV.  Noting that $ \hbar c/ R_{0} \approx 2 \times 10 ^{20} $MeV, we see that the negative gravitational energy (70) will cancel the $\hbar c/R_{0}$ term if $l \approx 3 \times 10 ^{-9}$ cm.  In view of Eq. (69), this also implies that short-range corrections to gravity would appear as long as $r < 10^{-9}$ cm.  On the other hand, if $R_{0}/\xi_{0} \approx 10^{2}$,  Eq. (64) yields $R_{0} \approx 10 ^{-29}$ cm.  Then, $\hbar c/R_{0} \approx 2 \times 10 ^{18}$ MeV and $E _{\textrm{grav}} \approx -2 \times 10 ^{31} l ^{2}$ MeV.  These results imply that cancellation of $\hbar c/R_{0}$ and $E _{\textrm{grav}}$ occurs for $l \approx 3 \times 10^{-7}$ cm.

Therefore, only an order of magnitude accuracy can be attached to estimates of electron mass from $E _{\textrm{tot}}$.  Nevertheless, the above cancellation of the $\hbar/R$ terms by negative gravitational energy is essential for the consistency of the model of Eq. (62).

\end{document}